%% file: Draft.tex
\documentclass[a4paper,11pt]{article}

\usepackage{jheppub} 

\makeatletter
\def\@fpheader{\newline}
\makeatother
\usepackage[T1]{fontenc} 

\input{AdditionalPreamble.tex}

\input{CustomCommands.tex}

\title{\texttt{RGESolver} :  a \texttt{C++} library to perform Renormalization Group evolution in the Standard Model Effective Theory}

\author[a,b]{S. Di Noi}
\author[c]{and L. Silvestrini}

\affiliation[a]{Dipartimento di Fisica e Astronomia ``G. Galilei'', Universit{\`a} degli Studi di Padova, Via Francesco Marzolo 8, I-35131 Padova, Italy}
\affiliation[b]{Istituto Nazionale di Fisica Nucleare, Sezione di Padova, Via Francesco Marzolo 8, I-35131 Padova, Italy}
\affiliation[c]{Istituto Nazionale di Fisica Nucleare, Sezione di Roma, Piazzale Aldo Moro 2, I-00185 Rome, Italy
}

\emailAdd{stefano.dinoi@phd.unipd.it}
\emailAdd{Luca.Silvestrini@roma1.infn.it}

\abstract{Renormalization group evolution above the electroweak scale
	is a crucial ingredient in the phenomenology of the Standard Model
	Effective Theory.  The \lstinline{RGESolver} open-source \lstinline{C++}~library
	performs the evolution at leading order for dimension-six operators
	in the most general flavour scenario (assuming lepton and baryon
	number conservation). Given its efficiency, \lstinline{RGESolver} can
	be used to include the effects of renormalization group evolution in
	extensive phenomenological analyses in the framework of the Standard
	Model Effective Theory.
}

\begin{document}

	
	
	
	\maketitle
	
	\section{Introduction}
	\label{sec:intro} 
	
	The Standard Model (SM) is one of the biggest scientific successes of
	our time: it describes three (weak, strong and electromagnetic) of the
	four currently known elementary interactions in nature, closing a long
	path started between the 19th and the 20th century. The particle
	content of the SM was completed in 2012 with the discovery of a
	Higgs-like boson with properties consistent with the SM within current
	experimental errors. In spite of its success,
	however, the SM leaves several phenomena unexplained. For example,
	neutrino masses, the origin of the baryon asymmetry in the universe
	and dark matter suggest the presence of physics beyond the SM. The
	absence of direct evidence of new particles at energies
	$\order{\mathrm{TeV}}$ allows us to parametrize the effects of
	possible heavy New Physics (NP), lying beyond the reach of the LHC for
	direct production, with an effective field theory, known as the
	Standard Model Effective Field Theory (SMEFT)
	\cite{Buchmuller:1985jz,dim6smeft}. The SMEFT Lagrangian contains the
	SM Lagrangian plus a complete set of independent higher-dimensional
	operators.  Working in the framework of the SMEFT makes it possible to
	search for NP through its virtual effects in a general,
	model-independent way.
	
	Going from the SM to the SMEFT entails dramatic phenomenological
	consequences, since the SM enjoys several accidental symmetries that
	are potentially broken by higher-dimensional operators, such as Baryon
	(B) and Lepton (L) number conservation, or the absence of tree-level
	Flavour Changing Neutral Currents (FCNC). Phenomenology requires the
	coefficients of those higher-dimensional operators that violate
	accidental symmetries of the SM to be tiny, implying in turn that any
	NP not too far from the electroweak (EW) scale must be at least
	approximately invariant under the accidental symmetries of the SM.
	However, while SM interactions will not generate B or L violation
	perturbatively if the corresponding operators in the SMEFT have
	vanishing coefficients, FCNC's will always be generated, even if NP is
	invariant under the full $\group{U(3)}^5$ flavour symmetry group of SM gauge
	interactions, due to the SM Yukawa couplings. Therefore, the flavour
	properties of the SMEFT Wilson coefficients must be specified at the
	NP scale, and then the coefficients must be evolved using
	Renormalization Group Equations (RGE's) down to the scale relevant for
	the processes of interest in order to compute the NP
	contributions. Conversely, a phenomenological bound on low-scale
	Wilson coefficients can be turned into a bound on the coefficients at
	the NP scale, allowing to extract information on the viable values of
	SMEFT coefficients and, hopefully, on the symmetries of the NP models
	of interest.
	
	Assuming B and L conservation but a general flavour structure, the
	SMEFT has 2499 independent operators, so that the full system of RGE's
	involves more than 2500 parameters. At Leading Order (LO), the renormalization group (RG)
	evolution is dictated by the Anomalous Dimension Matrices (ADM's) of
	the SMEFT operators, which in general receive contributions from
	scale-dependent gauge and Yukawa couplings, as well as from the Higgs
	self-coupling. Since the ADM contributions proportional to different
	couplings are in general non-commuting, an analytic resummation of
	logarithmic contributions cannot be achieved and the numerical
	solution of the full system of equations is the only possibility, in
	particular when the NP scale is much heavier than the EW one.
	\lstinline{RGESolver} is an open-source \lstinline{C++}~library that performs the RG
	evolution of the SMEFT Wilson coefficients in a fast and easy-to-use
	manner, as detailed below.
	\lstinline{RGESolver} will be also integrated in
	\lstinline{HEPfit}~\cite{hepfit}, a flexible open-source tool which, given the
	Standard Model or any of its extensions, allows to fit the model
	parameters to a given set of experimental observables and to obtain
	predictions for observables.
	
	This paper is organized as
	follows: we briefly introduce the theoretical framework of the SMEFT in
	Section~\ref{sec:smeft}, presenting the notation used in
	\lstinline{RGESolver}. In Section~\ref{sec:description} we
	describe in more detail the structure of the library, with a specific
	focus on the handling of the flavour structure and on the implementation of the numerical solution of the RGE's.
	Section~\ref{sec:usage} is devoted to describing the usage of
	\lstinline{RGESolver}: we discuss the installation procedure and we
	describe how to use the basic functionalities of the library, together with a list of its most important methods. We discuss the efficiency of the library and the comparison with \lstinline{DSixTools}~(\cite{dsixtools1,dsixtools2}) in Section~\ref{sec:efficiency}.
	We present our conclusions in Section~\ref{sec:conclusions}.
	
	\section{Theoretical framework: the SMEFT}
	\label{sec:smeft}
	
	As discussed above, the absence of new particles at energies at or
	above the TeV scale allows us to parametrize the effects of physics
	beyond the SM with a tower of dimension $\mathfrak{D}>4$, Lorentz and
	gauge-invariant operators \cite{Buchmuller:1985jz,dim6smeft}. The resulting
	effective field theory is the SMEFT. The SM gauge group is
	\begin{equation}
		\label{eq:SMgroup}
		\mathcal{G}_{\mathrm{SM}}\equiv\group{SU(3)}_{\mathrm{C}} \otimes\group{SU(2)}_{\mathrm{W}}\otimes\group{U(1)}_{\mathrm{Y}}.
	\end{equation}
	The three factors represent color, weak isospin and hypercharge gauge group. The quantum numbers of SM fields under $\mathcal{G}_{\mathrm{SM}}$ are listed in Table~\ref{tab:SMfields}.
	\begin{table}[h!]
		\centering
		\renewcommand{\arraystretch}{1.3} 
		\begin{tabular}{cc}
			\begin{tabular}[t]{cl}
				\hline
				Field & $\;\;\qm{\repr{R}_{\mathrm{C}}}{\repr{R}_{\mathrm{W}}}{\hy{}}$        \\ \hline
				$q^p$ & $\quad\qm{\repr{3}}{\repr{2}}{+\frac{1}{6}}$  \\
				$l^p$ & $\quad\qm{\repr{1}}{\repr{2}}{-\frac{1}{2}}$  \\
				$u^p$ & $\quad\qm{\repr{3}}{\repr{1}}{+\frac{2}{3}}$  \\
				$d^p$ & $\quad\qm{\repr{3}}{\repr{1}}{-\frac{1}{3}}$ \\
				$e^p$ & $\quad\qm{\repr{1}}{\repr{1}}{-1}$   
			\end{tabular} & \hspace{8 pt }
			\begin{tabular}[t]{cl}
				\hline
				Field & $\;\;\qm{\repr{R}_{\mathrm{C}}}{\repr{R}_{\mathrm{W}}}{\hy{}}$        \\ \hline
				$H$ & $\quad\qm{\repr{1}}{\repr{2}}{+\frac{1}{2}}$  \\
				$G^A_\mu$ & $\quad\qm{\repr{8}}{\repr{1}}{0}$  \\
				$W^I_\mu$ & $\quad\qm{\repr{1}}{\repr{3}}{0}$  \\
				$B_\mu$ & $\quad\qm{\repr{1}}{\repr{1}}{0}$  
			\end{tabular}
		\end{tabular}
		\caption{$\group{SU(3)}_{\mathrm{C}} \otimes\group{SU(2)}_{\mathrm{W}}\otimes\group{U(1)}_{\mathrm{Y}}$ quantum numbers of SM fields. $p$ is a flavour index. }
		\label{tab:SMfields}
	\end{table}
	The Lagrangian of the Standard Model is (following the notation used in Refs.~\cite{rge1,rge2,rge3})
	\begin{equation}
		\label{eq:SMlag}
		\begin{split}
			\lag_{\textrm{SM}} =&  -\frac{1}{4} G_{\mu \nu}^A G^{A\mu \nu}
			-\frac{1}{4} W_{\mu \nu}^I W^{I\mu \nu}
			-\frac{1}{4} B_{\mu \nu}B^{\mu \nu} \\
			&+ \sum_{\psi} \bar{\psi} i \slashed{D} \psi 
			+ (D_\mu H^\dagger) (D^\mu H) \\ &- \lambda \left(H^\dagger H - \frac{1}{2}v^2 \right)^2 \\
			&- \left[ H^\dagger \bar{d} \yuk{d} q +  \tilde{H}^\dagger \bar{u} \yuk{u} q 
			+  H^\dagger \bar{e} \yuk{e} l + \textrm{H.c.}   \right] .
		\end{split}
	\end{equation}
	With this notation, when spontaneous symmetry breaking occurs, the
	physical Higgs boson acquires a mass given by $m^2_H = 2 \lambda v^2$,
	with $v = (\sqrt{2} G_F)^{-1/2}\sim 246$ GeV.  The sum over $\psi$ is
	extended to all fermion fields in the SM, both left-handed ($q$, $l$)
	and right-handed ($u$, $d$, $e$). The gauge covariant derivative is
	$D_\mu = \partial_\mu$ $+ i g_1 \hy{} B_\mu$ $+i g_2 t^I W_\mu^I$$+i
	g_3 \gl{}{A}G_\mu^A$, with
	$\gl{}{A}$ and $t^I=\pauli{}{I}/2$\footnote{$\pauli{I}{}$,
		$I=1,2,3$, are the Pauli matrices: $\pauli{}{1} =
		\bigl( \begin{smallmatrix}0 & 1\\ 1 &
			0\end{smallmatrix}\bigr)$, $\pauli{}{2} =
		\bigl( \begin{smallmatrix}0 & -i\\ i &
			0\end{smallmatrix}\bigr)$, $\pauli{}{3} =
		\bigl( \begin{smallmatrix}1 & 0\\ 0 &
			-1\end{smallmatrix}\bigr)$.  } the generators in the fundamental
	representation of $\group{SU(3)}$ and $\group{SU(2)}$ respectively:
	\begin{equation}
		[\gl{}{A},\gl{}{B}] = i f^{ABC} \gl{}{C}, \quad
		[t^{I},t^{J}] = i \varepsilon^{IJK} t^K. 
	\end{equation}
	Finally, $\tilde{H}$ and $\tilde{X}$ (with $X =
	B,W^I,G^A$) are defined via the totally antisymmetric Levi-Civita
	symbol, with $\varepsilon_{12}=1$ and $\varepsilon_{0123}=1$:
	\begin{equation}
		\tilde{H_j} = \varepsilon_{jk} H^{* k}, \quad \tilde{X}^{\mu \nu} = \frac{1}{2} \varepsilon^{\mu \nu \rho \sigma} X_{\rho \sigma}.   
	\end{equation}
	
	Fermions appear in $n_g =
	3$ different generations, so every fermion field carries, together
	with the gauge indices, a generation (or flavour) index that runs from
	$1$ to
	$3$. We did not write explicitly neither of these indices in
	\eqref{eq:SMlag} for the sake of simplicity. Yukawa couplings
	$Y_\psi$ are thus complex matrices in flavour space.
	
	\begin{table}[H]
		\centering
		\scriptsize
		\renewcommand{\arraystretch}{1.3} 
		
		\begin{tabular}{ccc}
			\begin{tabular}[t]{l|c}
				\multicolumn{2}{c}{$1 : X^3$} \\ \hline
				{$\op{G}{}$}         & $f^{ABC}G_\mu ^{A \nu} G_\nu ^{B \rho}G_\rho ^{C \mu}$                    \\
				{$\op{\tilde{G}}{}$} & $f^{ABC} \tilde{G}_\mu ^{A \nu} G_\nu ^{B \rho}G_\rho ^{C \mu}$           \\
				{$\op{W}{}$}         & $\varepsilon^{IJK}W_\mu ^{I \nu} W_\nu ^{J \rho}W_\rho ^{K \mu}$          \\
				{$\op{\tilde{W}}{}$} & $\varepsilon^{IJK} \tilde{W}_\mu ^{I \nu} W_\nu ^{J \rho}W_\rho ^{K \mu}$ \\ [7 pt]
				\multicolumn{2}{c}{$2 : H^6$} \\ \hline	
				{$\op{H}{}$}         & $(H^\dagger H)^3$          \\
			\end{tabular} &  
			\begin{tabular}[t]{l|c}
				\multicolumn{2}{c}{$3 : H^4 D^2$} \\ \hline
				{$\op{H \Box}{}$}         & $(H^\dagger H) \Box(H^\dagger H) $          \\
				{$\op{HD}{}$} & $(H^\dagger D^\mu H)^* (H^\dagger D_\mu H) $ \\[7 pt]
				\multicolumn{2}{c}{$5 : \psi^2 H^3$} \\ \hline
				{$\op{eH}{}$} & $(H^\dagger H) (\bar{l}_p e_r H)$           \\
				{$\op{uH}{}$} & $(H^\dagger H) (\bar{q}_p u_r \tilde{H})$           \\
				{$\op{dH}{}$} & $(H^\dagger H) (\bar{q}_p d_r H)$       
			\end{tabular}  & 
			\begin{tabular}[t]{l|c}
				\multicolumn{2}{c}{$4 : X^2 H^2 $} \\ \hline
				{$\op{HG}{}$} & $ (H^\dagger H) G_{\mu \nu} ^A G^{A\mu \nu}$           \\
				{$\op{H\tilde{G}}{}$} & $ (H^\dagger H) \tilde{G}_{\mu \nu} ^A G^{A\mu \nu}$           \\
				{$\op{HW}{}$} & $ (H^\dagger H) W_{\mu \nu} ^I W^{I\mu \nu}$           \\
				{$\op{H\tilde{W}}{}$} & $ (H^\dagger H) \tilde{W}_{\mu \nu} ^I W^{I\mu \nu}$           \\
				{$\op{HB}{}$} & $ (H^\dagger H) B_{\mu \nu}  B^{\mu \nu}$           \\
				{$\op{H\tilde{B}}{}$} & $ (H^\dagger H) \tilde{B}_{\mu \nu}  B^{\mu \nu}$           \\
				{$\op{HWB}{}$} & $ (H^\dagger \pauli{}{I} H) W_{\mu \nu} ^ I  B^{\mu \nu}$           \\
				{$\op{H\tilde{W}B}{}$} & $ (H^\dagger \pauli{}{I} H) \tilde{W}_{\mu \nu} ^ I  B^{\mu \nu}$      
			\end{tabular} 
		\end{tabular}
		\par
		\vspace{10 pt }
		\begin{tabular}{cc}
			\begin{tabular}[t]{l|c}
				\multicolumn{2}{c}{$6 : \psi^2 X H$} \\ \hline
				{$\op{eW}{}$}         & $(\bar{l}_p \sigma^{\mu \nu} e_r)\pauli{}{I} H W_{\mu \nu}^I$                    \\
				{$\op{eB}{}$}         & $(\bar{l}_p \sigma^{\mu \nu} e_r) H B_{\mu \nu}$                    \\
				{$\op{uG}{}$}         & $(\bar{q}_p \gl{}{A} \sigma^{\mu \nu} u_r) \tilde{H} G_{\mu \nu}^A$                    \\
				{$\op{uW}{}$}         & $(\bar{q}_p \sigma^{\mu \nu} u_r)\pauli{}{I} \tilde{H} W_{\mu \nu}^I$                    \\
				{$\op{uB}{}$}         & $(\bar{q}_p \sigma^{\mu \nu} u_r) \tilde{H} B_{\mu \nu}$                   \\
				{$\op{dG}{}$}         & $(\bar{q}_p \gl{}{A} \sigma^{\mu \nu} d_r) H G_{\mu \nu}^A$                    \\
				{$\op{dW}{}$}         & $(\bar{q}_p \sigma^{\mu \nu} d_r)\pauli{}{I} H W_{\mu \nu}^I$                    \\
				{$\op{dB}{}$}         & $(\bar{q}_p \sigma^{\mu \nu} d_r) H B_{\mu \nu}$                     
			\end{tabular} &
			
			\begin{tabular}[t]{l|c}
				\multicolumn{2}{c}{$7 : \psi^2 H^2 D$} \\ \hline
				{$\op{Hl(1)}{}$}     & $(H^\dagger i\lrarrow{D} _\mu H)(\bar{l}_p \gamma^\mu l_r)$          \\
			{$\op{Hl(3)}{}$}      & $(H^\dagger i \lrarrow{D} ^I _\mu H)(\bar{l}_p \pauli{}{I}\gamma^\mu l_r)$          \\
			{$\op{He}{}$}     & $(H^\dagger i \lrarrow{D}_\mu  H)(\bar{e}_p \gamma^\mu e_r)$          \\
			{$\op{Hq(1)}{}$}     & $(H^\dagger i \lrarrow{D} _\mu H)(\bar{q}_p \gamma^\mu q_r)$          \\
			{$\op{Hq(3)}{}$}    & $(H^\dagger i \lrarrow{D}_\mu ^I H)(\bar{q}_p \pauli{}{I}\gamma^\mu q_r)$          \\
			{$\op{Hu}{}$}     & $(H^\dagger i \lrarrow{D} _\mu  H)(\bar{u}_p \gamma^\mu u_r)$          \\
			{$\op{Hd}{}$}     & $(H^\dagger i \lrarrow{D} _\mu  H)(\bar{d}_p \gamma^\mu d_r)$          \\
				{$\op{Hud}{}$}     & $(\tilde{H}^\dagger i D_\mu  H)(\bar{u}_p \gamma^\mu d_r)$          
			\end{tabular}  	
		\end{tabular}
		\par
		\vspace{10 pt}
		\begin{tabular}[t]{cc}
			\begin{tabular}[t]{l|c}
				\multicolumn{2}{c}{$8 : (\bar{L}L)(\bar{L}L)$} \\ \hline
				{$\op{ll}{}$}         & $(\bar{l}_p \gamma_\mu l_r)(\bar{l}_s\gamma^\mu l_t)$                   \\
				{$\op{qq(1)}{}$}         & $(\bar{q}_p \gamma_\mu q_r)(\bar{q}_s\gamma^\mu q_t)$                   \\
				{$\op{qq(3)}{}$}         & $(\bar{q}_p \gamma_\mu \pauli{}{I} q_r)(\bar{q}_s\gamma^\mu \pauli{}{I} q_t)$                   \\
				{$\op{lq(1)}{}$}         & $(\bar{l}_p \gamma_\mu l_r)(\bar{q}_s\gamma^\mu q_t)$                   \\
				{$\op{lq(3)}{}$}         & $(\bar{l}_p \gamma_\mu \pauli{}{I} l_r)(\bar{q}_s\gamma^\mu \pauli{}{I} q_t)$                   
				
			\end{tabular} &
			\begin{tabular}[t]{l|c}
				\multicolumn{2}{c}{$8 : (\bar{R}R)(\bar{R}R)$} \\ \hline
				{$\op{ee}{}$}         & $(\bar{e}_p \gamma_\mu e_r)(\bar{e}_s\gamma^\mu e_t)$                   \\
				{$\op{uu}{}$}         & $(\bar{u}_p \gamma_\mu u_r)(\bar{u}_s\gamma^\mu u_t)$                   \\
				{$\op{dd}{}$}         & $(\bar{d}_p \gamma_\mu d_r)(\bar{d}_s\gamma^\mu d_t)$                   \\
				{$\op{eu}{}$}         & $(\bar{e}_p \gamma_\mu e_r)(\bar{u}_s\gamma^\mu u_t)$                   \\
				{$\op{ed}{}$}         & $(\bar{e}_p \gamma_\mu e_r)(\bar{d}_s\gamma^\mu d_t)$                   \\
				{$\op{ud(1)}{}$}         & $(\bar{u}_p \gamma_\mu u_r)(\bar{d}_s\gamma^\mu d_t)$                   \\
				{$\op{ud(8)}{}$}         & $(\bar{u}_p \gamma_\mu \gl{}{A} u_r)(\bar{d}_s\gamma^\mu \gl{}{A} d_t)$          
			\end{tabular} 
		\end{tabular}
		\par
		\vspace{10 pt}
		\begin{tabular}[t]{cc}
			\begin{tabular}[t]{l|c}
				\multicolumn{2}{c}{$8 : (\bar{L}L)(\bar{R}R)$} \\ \hline
				{$\op{le}{}$}         & $(\bar{l}_p \gamma_\mu l_r)(\bar{e}_s\gamma^\mu e_t)$                   \\
				{$\op{lu}{}$}         & $(\bar{l}_p \gamma_\mu l_r)(\bar{u}_s\gamma^\mu u_t)$                   \\
				{$\op{ld}{}$}         & $(\bar{l}_p \gamma_\mu l_r)(\bar{d}_s\gamma^\mu d_t)$                   \\
				{$\op{qe}{}$}         & $(\bar{q}_p \gamma_\mu q_r)(\bar{e}_s\gamma^\mu e_t)$                   \\
				{$\op{qu(1)}{}$}         & $(\bar{q}_p \gamma_\mu q_r)(\bar{u}_s\gamma^\mu u_t)$                   \\
				{$\op{qu(8)}{}$}         & $(\bar{q}_p \gamma_\mu \gl{}{A} q_r)(\bar{u}_s\gamma^\mu \gl{}{A}  u_t)$                   \\
				{$\op{qd(1)}{}$}         & $(\bar{q}_p \gamma_\mu q_r)(\bar{d}_s\gamma^\mu d_t)$                   \\
				{$\op{qd(8)}{}$}         & $(\bar{q}_p \gamma_\mu \gl{}{A} q_r)(\bar{d}_s\gamma^\mu \gl{}{A}  d_t)$                   
			\end{tabular} & 
			\begin{tabular}[t]{l|c}
				\multicolumn{2}{c}{$8 : (\bar{L}R)(\bar{R}L)$} \\ \hline
				{$\op{ledq}{}$}         & $(\bar{l}_p e_r)(\bar{d}_s q_t)$   \\ [7 pt]             
				\multicolumn{2}{c}{$8 : (\bar{L}R)(\bar{L}R)$} \\ \hline
				{$\op{quqd(1)}{}$}         & $(\bar{q}_p^{j} u_r)\varepsilon_{jk}(\bar{q}_s^{k} d_t)$ \\  
				{$\op{quqd(8)}{}$}         & $(\bar{q}_p^{j} \gl{}{A} u_r)\varepsilon_{jk}(\bar{q}_s^{k}\gl{}{A} d_t)$               \\
				{$\op{lequ(1)}{}$}         & $(\bar{l}_p^{j} e_r)\varepsilon_{jk}(\bar{q}_s^{k} u_t)$ \\  
				{$\op{lequ(3)}{}$}          & $(\bar{l}_p^{j} \sigma_{\mu \nu} e_r)\varepsilon_{jk}(\bar{q}_s^{k} \sigma^{\mu \nu} u_t)$ 
			\end{tabular}	
		\end{tabular}%
		
		\normalsize
		\vspace{2 pt}
		\caption{The 59 independent dimension-six operators built from
			SM fields which conserve baryon and lepton number. $p,r,s,t$
			are fermion flavour indices, $j,k$ are indices in the
			fundamental representation of $\group{SU(2)}_{\mathrm{W}}$,
			$I,J,K$ ($A,B,C$) are indices in the adjoint representation
			of $\group{SU(2)}_{\mathrm{W}}$
			($\group{SU(3)}_{\mathrm{C}}$) and greek letters
			($\mu,\nu \dots$) are Lorentz indices. Contraction of
			indices in the fundamental representation of
			$\group{SU(3)}_{\mathrm{C}}$ is implicit.}
		\label{tab:SMEFToperators}
	\end{table}
	
	At dimension six a complete basis of independent and gauge invariant
	operators that conserve B and L is given by the so-called
	\textit{Warsaw basis}, defined in \cite{dim6smeft} and displayed in
	Table~\ref{tab:SMEFToperators}. Their independence means that no
	linear combination of them and their Hermitian conjugates vanishes due
	to the equations of motion (up to total derivatives). The equations of
	motion are used at $\order{1/\Lambda^2}$ level, so they can be
	derived from $\lag_{\mathrm{SM}}$ alone.
	
	The Warsaw basis consists of
	$59$ operators, for a total of $2499$ independent parameters in the
	generic flavour scenario (see Appendix A of Ref.~\cite{rge3}).  The
	operators are divided in $8$ classes, depending on their field
	content, which schematically are
	\begin{equation} \label{eq:SMEFToperatorclasses}
		\begin{tabular}{lll}
			$1: X^3,$       & $2:H^6,$        & $3: H^4 D^2,$      \\
			$4:X^2 H^2,$ & $ 5:\psi^2 H^3,$ & $6: \psi^2 HX,$ \\
			$7:\psi^2 H^2 D,$ & $8: \psi^4,$ & $ $
		\end{tabular}
	\end{equation}
	where $X$ stands for a gauge field-strength tensor (or its dual),
	$\psi$ for a fermion field, $D$ for a derivative and $H$ is the Higgs
	field.

\begin{figure}[h!]
		\begin{subfigure}[t]{.49\linewidth}
			\centering
			\begin{tikzpicture}
				\begin{feynman}
					\vertex  (u1) {\( H \)}  ;
					\vertex [square dot, scale = 1.3, right =  40 pt of u1 ] (O) {};
					\vertex [right = 40 pt of O] (u2) {\( H \)};	
					\vertex  [below left = of O] (u4) {\( H \)}  ;
					\vertex  [below right =  40 pt of O] (u3) {\( H \)}  ;
					\vertex  [above = 28.3  pt of O,square dot, scale = 0.01] (V) {}  ;
					\diagram* {
						(u1) -- [scalar] (O) -- [scalar] (u2),	
						(O) --[min distance=0.01 pt, scalar,half left] (V) 
						-- [min distance=0.01 pt, scalar,half left] (O),
						(u3) -- [scalar] (O) -- [scalar] (u4),	
					};
				\end{feynman}
			\end{tikzpicture}
			\caption{SMEFT diagram contributing to the $\beta$ function of $\lambda$ with terms $\order{m^2_H \coeff{H}{}} $.}
			\label{diag:smefttosmrunning1}
		\end{subfigure}
		\hfill
		\begin{subfigure}[t]{.49\linewidth}
			\centering
			\begin{tikzpicture}
				\begin{feynman}
					\vertex  (u1) {\( H \)}  ;
					\vertex [square dot, scale = 1.3, below right =  40 pt of u1 ] (O) {};
					\vertex  [right = 40 pt of O, dot, scale = 1] (V) {}  ;	
					\vertex [above right =  40 pt of V] (u2) {\( H \)};
					\vertex  [below left =  40 pt of O] (u4) {\( H \)}  ;
					\vertex  [below right =  40 pt of V] (u3) {\( H \)}  ;
					\diagram* {
						(u1) -- [scalar] (O) -- [scalar] (u4),	
						(O) --[scalar, half left] (V) -- [scalar, half left] (O),
						(u3) -- [scalar] (V) -- [scalar] (u2),	
					};
				\end{feynman}
			\end{tikzpicture}
			\caption{SMEFT diagram contributing to the $\beta$ function of $\lambda$ with terms $\order{ \lambda m^2_H \coeff{HD}{}} $,$\order{\lambda m^2_H \coeff{H \Box}{}}$.}
			\label{diag:smefttosmrunning2}
		\end{subfigure}
		\caption{Two SMEFT diagrams that contribute to the running of the Higgs quartic coupling $\lambda$. The solid square denotes a SMEFT vertex and the dot denotes a SM vertex. }
		\label{fig:smefttosmrunning}
	\end{figure}
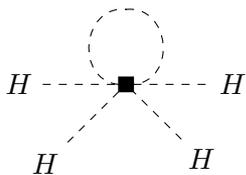
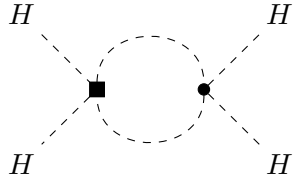

	To conclude the discussion about the SMEFT we note that higher
	dimensional operators contribute to observables not only through their
	matrix element, but also through their contribution to the running of
	SM couplings.  An example is given by the diagrams in
	Figure~\ref{fig:smefttosmrunning}. Coherently with the power counting,
	their effects in the running of $\lambda$ are suppressed by a factor
	of $m^2_H/\Lambda^2$.  The contributions to the running of the SM parameters due to SMEFT operators are available in Ref.~\cite{rge1}.
	
	\subsection{Flavour basis and back-rotation}
	\label{sec:backrotation}
	In the SM a unitary transformation for each fermion field in flavour space,
	\begin{equation}
		\label{eq:flavourrotation}
		\psi = R_\psi^\dagger \psi^\prime, \qquad \psi = q,l,u,d,e,
	\end{equation}
	does not alter the structure of the Lagrangian, provided the Yukawa
	matrices are redefined as
	\begin{equation}
		\label{eq:yukawaredefinition}
		\begin{cases}
			\yuk{u} = R_u^\dagger \yuk{u}^\prime R_q,\\
			\yuk{d} =  R_d^\dagger \yuk{d}^\prime R_q,\\
			\yuk{e} = R_e^\dagger \yuk{e}^\prime R_l.
		\end{cases}
	\end{equation}
	
	This freedom allows to diagonalize simultaneously two out of the three
	matrices ($\yuk{e}$ and one between $\yuk{u}$ and
	$\yuk{d}$).\footnote{The situation changes after the spontaneous
		symmetry breaking in the SM: fermions acquire a mass matrix proportional
		to the corresponding Yukawa matrix, $m_\psi^{pr}=$
		$v \yuk{\psi}^{pr}/\sqrt{2}$, $\psi=$$u$, $d$,
		$e$. All the mass matrices can be simultaneously diagonalized
		because the up and down component of the iso-doublet
		$q$ can be rotated indipendently.} Denoting with
	${\hat{\yuk{}}}$ a diagonal Yukawa matrix we can define two reference
	bases: the \textit{down basis} where ${\yuk{e}}=
	$${\hat{\yuk{e}}}$, ${\yuk{d}}= $${\hat{\yuk{d}}}$,
	${\yuk{u}}=
	$${\hat{\yuk{u}}} V$ and the \textit{up basis} where
	${\yuk{e}}= $${\hat{\yuk{e}}}$, ${\yuk{u}}=
	$${\hat{\yuk{u}}}$, ${\yuk{d}}= $${\hat{\yuk{d}}} V^\dagger$, with
	$V$ a unitary matrix. In the SM,
	$V$ is the Cabibbo-Kobayashi-Maskawa matrix. In the SMEFT this is not
	true, since the mass matrices are not simply proportional to the
	Yukawa couplings due to the contributions from dimension-six
	operators. With a slight abuse of notation, in the following we still
	refer to the matrix
	$V$ as the CKM matrix. It is worth noticing that the matrix
	$V$ is not the one appearing in the interaction vertices between
	quarks and
	$W^{\pm}$ bosons, again due to effects from higher dimensional
	operators. A more detailed discussion is available in Section~2 of
	Ref.~\cite{jenkins2018low}.
	
	Notice that the up and down bases are not stable under
	renormalization group evolution, even in the SM. For example, the SM
	$\beta$ function for
	$\yuk{d}$ (available in Ref.~\cite{rgesm}) contains a term
	proportional to $\yuk{d}(\yuk{d}^\dagger \yuk{d}-\yuk{u}^\dagger
	\yuk{u}
	)$, which is always non-diagonal in flavour space. Starting from a
	diagonal $\yuk{d}$ at the scale
	$\mu_{\mathrm{I}}$ (down basis) leads to a non-diagonal
	$\yuk{d}$ at a different scale
	$\mu_{\mathrm{F}}$. To go back into the down basis, a further flavour
	rotation is thus required. Clearly, the same holds for the up
	basis.\newline In the SMEFT the flavour transformations in
	Equation~\eqref{eq:flavourrotation} imply not only a redefinition of the
	Yukawa couplings, but also a rotation of the Wilson coefficients. For
	example, the coefficient $\coeff{dH}{}$ of operator
	$\op{dH}{}$ is redefined as $\coeff{dH}{}= R_q^\dagger \coeff{dH}{\prime}
	R_d$. \newline It should be noticed that the flavour rotations are not
	uniquely determined by the singular value decomposition. The three
	Yukawa matrices are diagonalized via six unitary matrices:
	\begin{equation}
		\label{eq:yukawadiagonalization}
		\yuk{\psi}= U_\psi {\hat{\yuk{\psi}}} V^{\dagger}_\psi, \quad \psi = u,d,e.
	\end{equation}
	If
	$U_\psi,\,V_\psi$ satisfy the
	relation~(\ref{eq:yukawadiagonalization}), also $U_\psi
	\phi_\psi,\,V_\psi \phi_\psi$ with $\phi_\psi =
	\textrm{diag}\left(e^{\alpha_1^\psi},e^{\alpha_2^\psi},e^{\alpha_3^\psi}\right)$
	satisfy it. The phase matrices $\phi_u,\,
	\phi_d$ are determined unambiguosly once the phase convention for the
	CKM matrix is chosen, while, in the absence of right-handed neutrinos,
	the phase of the lepton matrices cannot be determined. A possibility
	to fix the phases would be using one of the SMEFT coefficients that
	involve leptons, but this would not be a general solution. In fact,
	the user may want to set that coefficient to
	$0$, frustrating the choice of phase convention. \newline To overcome
	this problem, we choose $\phi_e = \textrm{diag}\left( e^{-i \arg
		\left[ \left(U_e\right)_{11}\right]}, e^{-i \arg \left[
		\left(U_e\right)_{22}\right]}, e^{-i \arg \left[
		\left(U_e\right)_{33}\right]}\right)$.  This choice ensures that
	an evolution between two energy scales close to each other produces a
	small change in the coefficients. The user is then free to perform the
	rotation in the lepton sector choosing any other
	convention. \\
	To go into the up basis, the rotation matrix
	$R_q$ in~(\ref{eq:flavourrotation}) must be set equal to
	$V_u\phi_u$. To go into the down basis one must instead set $R_q =
	V_d\phi_d$. Obviously, in both bases one sets
	$R_u=U_u\phi_u$, $R_d=U_d\phi_u$, $R_l=V_e\phi_e$ and
	$R_e=U_e\phi_e$.

	\subsection{Solution of the renormalization group equations}
	\label{subsec:resolution}
	
	The RGE for a parameter $x_i$, that in this context is a SMEFT
	coefficient $\coeff{i}{}$ or a SM parameter, is
	\begin{equation}
		\label{eq:betafunction}
		\mu \frac{d x_i}{d \mu} \equiv \beta_{i}\left( \{x_j\}\right),
	\end{equation} 
	that defines the $\beta$-function for the parameter. In general, when the theory has multiple parameters, the $\beta$-function for $x_i$ can contain terms proportional to integer powers of $x_i$, but also terms proportional to other parameters $x_j$, $i \ne j$. 
	In this article we always consider the $\beta$-functions at one-loop level.
	
	The $\beta$-functions of SMEFT coefficients (as well as SMEFT
	contributions to the running of SM parameters) are available in Refs.~\cite{rge1,rge2,rge3}, the SM $\beta$-functions of gauge
	couplings can be found for example in Ref.~\cite{srednicki} and the
	$\beta$-functions of Yukawa couplings, $m_H^2$ and $\lambda$ in
	Ref.~\cite{rgesm}.\footnote{This reference uses a different convention for the Higgs sector and for $g_1$. The Higgs mass is $m^2_H =$
		$- m^2/2$ and the replacements $g_1 \to g_1\sqrt{5/3}$,
		$\lambda \to 2 \lambda$ are required to convert their results into
		our convention.}
	
	The $\beta$-function for SMEFT coefficients must be linear in the
	coefficients themselves by power counting. In fact, the product of two
	dimension-six SMEFT coefficients would be suppressed by a factor of
	$1/\Lambda^4$, negligible at $\order{m^2_H/\Lambda^2}$. This allows
	to write the RGE's for the $\coeff{i}{}$s introducing the
	\textit{anomalous dimension matrix} $\adm$ (ADM),
	\begin{equation}
		\label{eq:anomalousdimension}
		\mu \frac{ d \coeff{i}{}}{d \mu} = \frac{1}{16 \pi^2}\adm_{ij}\coeff{j}{}.
	\end{equation}
	
	The indices $i$, $j$ in the previous equation run on the whole set of
	independent coefficients: $\adm$ is a square $2499 \times 2499$
	matrix.  Clearly, the
	$\beta$-function for the $i$-th coefficient is
	$\beta_{{\coeff{i}{}}}=$ $\adm_{ij}\coeff{j}{}/(16 \pi^2)$.
	
	A simple approximation consists in neglecting the $\mu$ dependence of
	the right-hand side of Equation~\eqref{eq:anomalousdimension} and taking only
	linear order terms in
	$\ln \left( \mu_{\textrm{F}} / \mu_{\textrm{I}}\right)$, yielding
	\begin{equation}
		\label{eq:approximate}
		\coeff{i}{}(\mu_{\textrm{F}}) = \coeff{i}{}(\mu_{\textrm{I}}) + 
		\adm_{ij}(\mu_{\textrm{I}}) \coeff{j}{}(\mu_{\textrm{I}})  \frac{\ln{\left( \mu_{\textrm{F}} / \mu_{\textrm{I}}\right)}}{16 \pi^2}.
	\end{equation}
	While this drastic simplification may be appropriate for cases in
	which the two energy scales are not too different, so that the second
	term in Equation~\eqref{eq:approximate} is a small correction, in general
	the log on the right-hand side will become large, calling for a
	resummation of log-enhanced terms.
	
	To provide a more accurate solution of the system of first-order
	differential equations in Equation~\eqref{eq:anomalousdimension} is a
	non-trivial task since the anomalous dimension matrix depends on $\mu$
	through the SM couplings, the running of which is in turn influenced by
	SMEFT operators. The anomalous dimension matrix can be written as
	\begin{equation}
		\label{eq:anomalousdimension2}
		\adm_{ij} (\mu)= g_1^2 (\mu) \adm_{ij} ^{(g_1^2)} + g_2^2(\mu) \adm_{ij} ^{(g_2^2)} + \ldots, 
	\end{equation} 
	where the matrices on the right-hand side do not depend on the
	renormalization scale. In general the $\adm^{(g_i)}$ matrices cannot
	be diagonalized simultaneously: an exact analytic solution of the
	system is then impossible. One could proceed with a hybrid scheme as
	commonly done in the Effective Theory for Weak interactions (WET)
	below the EW scale, where QCD and QED corrections are implemented by
	resumming the logs generated by the strong coupling only. However, in
	the SMEFT the top Yukawa coupling is also large, calling for the
	resummation of both strong and Yukawa contributions.
	
	Thus, a numerical approach is required to obtain precise results in
	the general case.\footnote{An approximate resummation of strong and
		Yukawa contributions might be obtained by assuming the
		$y_t^2/\alpha_s$ ratio to be scale-invariant, as suggested in Ref.~\cite{Buras:2018gto}.}  More details about the implementation
	of this method are given in Section~\ref{subsec:numericalresolution}.
	
	
	\section{Description of the library}
	\label{sec:description}
	\lstinline{RGESolver} is implemented in \lstinline{C++}.  We use the GNU Scientific
	Library for the integration of the RGE's (more information about the
	numerical methods used are given below). The whole code (including
	input \& output) handles separately real and imaginary parts of
	the parameters.
	
	\subsection{Flavour structure}
	\label{subsec:flavour}
	Operators (or equivalently their coefficients) in the SMEFT can be
	divided in different categories according to their symmetry properties
	as in Table~\ref{tab:WC},\footnote{It is worth noticing that there exists a symmetry class specific for the operator $\op{ee}{}$, since
		it has an additional symmetry following from $e^p$ being
		a singlet under
		$\group{SU(3)}_{\mathrm{C}} \otimes \group{SU(2)}_{\mathrm{W}}$ (see Ref.~\cite{rge3}).} where we partially modify the notation used in
	Table~14 of Ref.~\cite{dsixtools2}.
	
	\begin{table}[h!]
		\centering
		\renewcommand{\arraystretch}{1.3} 
		\begin{tabular}[t]{cl}
			\hline
			Category & Symmetry properties        \\ \hline
			$0$ & 0F scalar object  \\
			$1$ & 2F generic real matrix  \\
			2R& 2F Hermitian matrix (Re)  \\
			2I& 2F Hermitian matrix (Im)  \\
			5 & 4F generic real object \\
			6R & 4F two identical $\bar{\psi} \psi$ currents (Re) \\
			6I & 4F two identical $\bar{\psi} \psi$ currents (Im) \\        
			7R & 4F two independent $\bar{\psi} \psi$ currents (Re) \\
			7I & 4F two independent $\bar{\psi} \psi$ currents (Im) \\ 
			8R & $\coeff{ee}{}$ (Re) \\
			8I & $\coeff{ee}{}$ (Im) \\
		\end{tabular}
		\caption{Symmetry categories for operators in the SMEFT. nF indicates the number of flavour indices for each category.}
		\label{tab:WC}
	\end{table}
	
	We work splitting real and imaginary part for each $\coeff{i}{}$.
	For example, the real part of the coefficient of an Hermitian 2F operator is a
	(real) symmetric matrix, while the imaginary part is a (real)
	antisymmetric matrix. Also the real and imaginary part of symmetric 4F
	operators are in different categories. This is not the case for
	non-Hermitian operators: the real and imaginary parts of their coefficient do not have
	any constraints, and they belong to categories 1 (2F) and 5
	(4F). Clearly it is convenient to store and evolve only the
	independent parameters, but for ease of use it is convenient to be
	able to access the coefficients with any combination of indices, not
	only the independent ones. We achieve this flexibility by defining specific
	symmetry-aware getters and setters for each category.
	For example, setting $\mathrm{Re}[\coeff{Hl(1)}{1,2}]=0.5$ via
	\lstinline{S.SetCoefficient}\lstinline{("CHl1R",0.5,1,2)} (where \lstinline{S}
	is an instance of the \lstinline{RGESolver} class) will lead to
	\lstinline{S.GetCoefficient}\lstinline{("CHl1R",2,1)} returning the value
	\lstinline{0.5} due to Hermiticity.
	
	\subsection{Numerical implementation of renormalization group evolution}
	\label{subsec:numericalresolution}
	As already mentioned, we use the routines provided by the GNU
	Scientific Library (\lstinline{GSL}, \cite{gsl}) to perform the
	integration of the RGEs.  Let $\vec{y}$ be the vector that contains
	all the independent parameters of the SMEFT.  Introducing
	$t = \ln ( \mu / \textrm{GeV})$, the system in
	Equation~\eqref{eq:betafunction} can be rewritten as
	\begin{equation}
		\label{eq:num1}
		\frac{d \vec{y}}{d t} = \vec{\mathrm{f}}(\vec{y}), 
	\end{equation}
	where the right-hand side of the equation depends on $t$ only through
	$\vec{y}$.  When all the symmetries of the operators are considered,
	the coefficients of operators at dimension-six level are completely
	determined through $2499$ independent real parameters.  The three
	gauge couplings $g_1$, $g_2$, $g_3$, the Higgs quartic coupling
	$\lambda$ and the mass $m_H$ of the Higgs boson are real scalars.
	Yukawa couplings $\yuk{u}$, $\yuk{d}$ and $\yuk{e}$ are $3\times 3$ complex matrices, each of which has $18$ parameters.\footnote{In
		the Yukawa sector not all of the $54$ real parameters are
		observable: only $13$ parameters from this sector are independent
		($9$ fermion masses plus the $3$ angles and one phase in the CKM
		matrix). However, above the electroweak scale the $\beta$ functions
		are naturally expressed in terms of the Yukawa matrices rather than
		of the fermion masses and the CKM matrix. Therefore, for the purpose
		of computing RG evolution it is more convenient to consider general
		Yukawa matrices.} This yelds $59$ further parameters, raising the
	total number to $2558$: this is the dimension of $\vec{y}$.  The
	system in Equation~\eqref{eq:betafunction} is thus a $2558$-dimensional
	system of first order coupled differential equations. 
	We use an adaptive stepsize routine to solve the system: the stepsize
	$h$ is changed throughout the integration in order to match the estimated
	local error $E_i$ with a user-defined error level, to obtain the
	maximum efficiency. The desired error level $D_i$ for each component
	is determined through four parameters $a_y$, $a_{dydt}$,
	$\epsilon_{\textrm{abs}}$ and $\epsilon_{\textrm{rel}}$ with the
	expression
	\begin{equation}
		\label{eq:num2}
		D_i = \epsilon_{\textrm{abs}} + \epsilon_{\textrm{rel}} \left( a_y \lvert{y_i}\rvert + h a_{dydt} \left\vert \frac{d y_i}{d t} \right\vert \right)
	\end{equation}
	The second term in brackets in the previous expression ensures a
	correct estimation also for situations where some components $y_i$ are
	very close to zero. The error is determined not only from the value
	$y_i$ but also from its increment. In particular, we use
	$a_y=$$a_{dydt}=1$.
	
	$E_i$ can be estimated with the \textit{step-doubling} technique, that
	consists in advancing the solution from $t$ to
	$t+2h$ in two different ways (performing two steps of length
	$h$ or one step of length
	$2h$) and taking the error as the difference between the two. A more
	sofisticated and efficient error estimate is possible for the
	\textit{embedded} integration methods, where the same evaluations of
	the function
	$\vec{\textrm{f}}$ are used to compute two different values of the
	solution at
	$t+h$. The error is taken as the difference between the two values
	(see Section~16.2 of Ref.~\cite{numericalrecipes}).
	
	If $E_i$ exceeds the desired error $D_i$ by more than $10 \%$ for any
	component, the stepsize is reduced according to
	\begin{equation}
		\label{eq:num3}
		h \to h \times 0.9 \times \left( \frac{E}{D}\right)^{-\frac{1}{q}}\,,
	\end{equation}
	where $0.9$ is a safety factor (the error can only be estimated, not
	accurately determined), $q$ is the consistency order of the method
	(\textit{i.e.} the local error scales as $h^{q+1}$) and
	$E/D = \max_i \left( E_i / D_i \right)$ is the maximum ratio of
	estimated and desired error among the components.
	
	If,
	instead, the estimated error is lower than the desired one (precisely,
	when $E/D$ $<50 \%$) the stepsize is increased according to
	\begin{equation}
		\label{eq:num4}
		h \to h \times 0.9 \times \left( \frac{E}{D}\right)^{-\frac{1}{q+1}}.
	\end{equation}
	To avoid uncontrolled changes in the stepsize, the overall scaling
	factor is limited to the range (1/5, 5).  In \lstinline{RGESolver} the
	explicit embedded Runge-Kutta-Fehlberg method is used (with initial
	stepsize
	$h=\ln (\mu/\Lambda)/100$), for
	which $q=4$.
	
	It should be stressed that Equation~\eqref{eq:num2}
	refers to a local error: there is no simple relation that connects
	the four parameters to an estimate of the global error affecting the
	final result. We tested the accuracy of the adaptive stepsize
	integration comparing it with a fixed-step integration with an high
	number of steps. Using $\mu_{\textrm{I}}= \Lambda = 1000\,$
	TeV and $\mu_{\textrm{F}}=250$ GeV as energy scales,
	$\epsilon_{\textrm{abs}}=10^{-16}$ and
	$\epsilon_{\textrm{rel}}=10^{-4}$ for the adaptive integration and
	$N_{\textrm{FS}}=1000$ steps for the fixed stepsize integration we
	obtained percentual differences $\lesssim 10^{-5}$.
	The initial conditions for SM parameters at the high energy scale are
	obtained evolving them from $\mu \sim v$ to $\Lambda$ with the pure
	SM $\beta$-functions (namely neglecting SMEFT contributions). The
	initial conditions for SMEFT coefficients are
	$\order{1/\Lambda^2}$, as prescribed by the power
	counting.

	\section{Using \texttt{RGESolver}}
	\label{sec:usage}
	
	\subsection{Installation}
	\label{subsec:install}
	\lstinline{RGESolver} is a free software released under the GNU General Public License. The download can be performed directly from the command line, typing in the terminal:
\begin{lstlisting}[language=bash]
git clone	https://github.com/silvest/RGESolver --recursive
\end{lstlisting}
More details can be found on the \href{https://github.com/silvest/RGESolver}{dedicated \lstinline{GitHub} webpage}. The extended documentation is also available \href{https://silvest.github.io/RGESolver/annotated.html}{here}.
	\subsubsection*{Dependencies}
	The installation of \lstinline{RGESolver} requires the availability of
	\lstinline{CMake} in the system (version \lstinline{3.1} or greater). A
	description of \lstinline{CMake} and the instructions for its
	installation can be found in the
	\href{https://cmake.org/}{\lstinline{CMake} website}.  We list below the
	dependencies that need to be satisfied to succesfully install
	\lstinline{RGESolver}:
	\begin{itemize}
		\item \lstinline{GSL}: The GNU Scientific Library (\lstinline{GSL}) is a \lstinline{C} library for numerical computations. More details can be found on the \href{https://www.gnu.org/software/gsl/}{\lstinline{GSL} website}.
		\item \lstinline{BOOST}: \lstinline{BOOST} is a set of libraries for
		the \lstinline{C++}~programming language. \lstinline{RGESolver} requires
		only the \lstinline{BOOST} headers, not the full libraries,
		thus a header-only installation is sufficient. More details
		can be found on the \href{https://www.boost.org/}{\lstinline{BOOST} website}.
		\item \lstinline{C++11}: a \lstinline{C++} compiler supporting at
		least the \lstinline{C++11} standard.
	\end{itemize}
	
	If all dependencies are satisfied, the installation procedure is
	completed typing in a terminal session in the downloaded
	\lstinline{RGESolver} directory:
	\begin{lstlisting}
		mkdir build && cd build
		cmake .. <options>
		cmake --build .
		cmake --install .
	\end{lstlisting}
	The available options are: 
	\begin{itemize}
		\item \lstinline{-DLOCAL_INSTALL=ON}: to install \lstinline{RGESolver} in the directory \lstinline{build/install} (default: \lstinline{OFF}).
		\item \lstinline{-DCMAKE_INSTALL_PREFIX=<RGESolver installation directory>}: the directory in which \lstinline{RGESolver}	will be installed (default: \lstinline{/usr/local}). This variable cannot be modified when \lstinline{-DLOCAL_INSTALL=ON} is set.

		\item \lstinline{-DDEBUG_MODE=ON}: to enable the debug mode (default: \lstinline{OFF}).
		
		\item \lstinline{-DBOOST_INCLUDE_DIR=<boost custom include path>/boost/}: \lstinline{CMake} checks for \lstinline{BOOST} headers availability in the system and fails if they
		are not installed. Thus, if \lstinline{BOOST} is not installed
		in the predefined search path, the user can specify where it is with this option. The path must end with
		the \lstinline{boost/} directory which contains the headers.
		
		\item \lstinline{-DGSL_CONFIG_DIR=<path to gsl-config>}: \lstinline{RGESolver} uses \lstinline{gsl-config} to get the \lstinline{GSL} parameters. If this is not in the predefined search path, the user can specify it with this option.
		
	\end{itemize}
	If no \lstinline{<options>} are specified, the default installation will be performed. Note that, depending on the setting of installation prefix, root privileges may be required (thus \lstinline{cmake -}\lstinline{-install . } should be replaced with \lstinline{sudo cmake -}\lstinline{-install .}).\newline 
	\lstinline{RGESolver} can be uninstalled by typing in the build directory of the \lstinline{RGESolver} library the command \lstinline{(sudo)} \lstinline{cmake -}\lstinline{-build . -}\lstinline{-target uninstall}.

	\subsection{Class methods}
	\label{subsec:methods}
	Here we briefly describe the most important methods available in \lstinline{RGESolver}. The detailed documentation is available on \href{https://silvest.github.io/RGESolver/annotated.html}{\lstinline{GitHub}}. 
	\subsubsection*{Evolution}
	\begin{itemize}
		
		\item \method{void Evolve(std::string method, double muI,
			double muF)}: performs the one-loop
		renormalization group evolution from $\mu = $ \lstinline{muI} to
		$\mu = $ \lstinline{muF} (where \lstinline{muI} and \lstinline{muF}
		are expressed in GeV). The current values of the SMEFT
		parameters are taken as initial condition at
		$\mu = $ \lstinline{muI}. After the evolution, they are updated
		with the new values at $\mu = $ \lstinline{muF}. The available
		methods for the solution of the RGE's are
		\lstinline{"Approximate"} and \lstinline{"Numeric"}. The first
		method is faster than the latter, but less
		accurate, as explained in Section~\ref{subsec:resolution}.
		
		\item \method{void EvolveSMOnly(std::string method, double muI, double muF)}: same as \lstinline{Evolve}, but only for the SM parameters. 
		The user should use this method instead of \lstinline{Evolve} when 
		interested in pure SM running. Using this function is the same of 
		using \lstinline{Evolve} with all the SMEFT coefficients set to 0, but it is faster since 
		it computes only the evolution for the SM parameters.
		\item \method{void EvolveToBasis(std::string method, double muI,
			double muF, std::string basis)}: same as \lstinline{Evolve}, but performs in addition 
		the flavour back-rotation described in Section~\ref{sec:backrotation} to go into the selected basis (\lstinline{"UP"} or \lstinline{"DOWN"}). After the evolution, the CKM matrix is computed.
		
		\item \method{void GenerateSMInitialConditions(...)}:
		generates the initial conditions for Standard Model
		parameters ($g_1, \,g_2, \,g_3, \, \lambda, \, m_h^2,$
		$\yuk{u},\,\yuk{d},\,\yuk{e}$) at the scale \lstinline{muFin}
		(in GeV), using one-loop pure SM beta functions. At such
		scale, the CKM matrix is computed.\newline The generation
		can be done starting from user-defined low-energy input or
		using the default values summarized in
		Table~\ref{tab:SMinput} (the arguments to be passed to the
		function are different in the two cases, see the
		documentation for the details and
		Subsection~\ref{subsec:example} for an example). In case of
		user-defined input, this method should be used with usual
		fermion hierarchy (smallest mass for the 1st generation and
		greatest mass for the 3rd without mass degeneracy for all
		up and down quarks and for charged leptons).
	\end{itemize}
	
	\begin{table}[h!]
		\centering
		\caption{SM parameters used by default to generate SM initial conditions at an arbitrary scale. The scale at which these parameters are given is $\mu = 173.6$ GeV. We follow \href{http://www.utfit.org/UTfit}{UTfit} for what concerns the conventions for the CKM matrix.}
		\label{tab:SMinput}
		\begin{tabular}[t]{c c} 
			\begin{tabular}{c|c}
				Parameter                     & Value        \\ \hline
				$g_1$                  & $0.3573$     \\
				$g_2$                   & $0.6511$   \\
				$g_3$    & $1.161$    \\
				$\lambda$           & $0.1297$\\
				$m_h^2$ $[$GeV$^2]$                          & $ 15650$ \\
				$\sin \theta_{12}$         & $0.2252$      \\
				$\sin \theta_{13}$          & $0.003675$     \\
				$\sin \theta_{23}$           & $0.0420$      \\
				$\delta$ $[$rad$]$                      & $1.1676$  
			\end{tabular} & \hfill
			\begin{tabular}{c|c}
				Parameter                     & Value [GeV]     \\ \hline
				$m_u$                  & $0.0012$     \\
				$m_c$                  & $0.640$     \\
				$m_t$                  & $162.0$     \\
				$m_d$                  & $0.0027$     \\
				$m_s$                  & $0.052$     \\
				$m_b$                  & $2.75	$     \\
				$m_e$                  & $0.000511$     \\
				$m_\mu$                  & $0.1057$     \\
				$m_\tau$                  & $ 1.776$     
			\end{tabular}
		\end{tabular}
		
	\end{table}
	
	\subsubsection*{Input \& Output}
	
	\begin{table}[h!]
		
		\scriptsize
		\renewcommand{\arraystretch}{1.3} 
		
		\begin{minipage}[t]{.55\linewidth}
			\centering
			\caption{Standard Model parameters in \texttt{RGESolver}. The labels in the left column must be used with the \texttt{GetCoefficient} method. The ones in the right column must be used with the \texttt{GetCKMAngle} method.}
			\label{tab:parnamesSM}
			\begin{tabular}[t]{cc} 
				\begin{tabular}[t]{c|c} 
					Coefficient     & \multicolumn{1}{c}{Name}      \\ \hline 
					$g_1$         & \texttt{g1}      \\
					$g_2$         & \texttt{g2}      \\
					$g_3$         & \texttt{g3}      \\
					$\lambda$         & \texttt{lambda}      \\
					$m_h^2$ $[\mathrm{GeV}^2]$  & \texttt{mh2}      \\
					$\mathrm{Re}(\mathcal{Y}_u)$         & \texttt{YuR}      \\
					$\mathrm{Im}(\mathcal{Y}_u)$         & \texttt{YuI}      \\
					$\mathrm{Re}(\mathcal{Y}_d)$         & \texttt{YdR}      \\
					$\mathrm{Im}(\mathcal{Y}_d)$         & \texttt{YdI}      \\
					$\mathrm{Re}(\mathcal{Y}_e)$         & \texttt{YeR}      \\
					$\mathrm{Im}(\mathcal{Y}_e)$         & \texttt{YeI}      
				\end{tabular} & 
				\begin{tabular}[t]{c|c} 
					Coefficient     & \multicolumn{1}{c}{Name}      \\ \hline 
					$\sin \theta_{12}$         & \texttt{s12}      \\
					$\sin \theta_{13}$         & \texttt{s13}      \\	
					$\sin \theta_{23}$         & \texttt{s23}     
				\end{tabular}
			\end{tabular}
		\end{minipage}%
		\hfill
		\begin{minipage}[t]{.42\linewidth}
			\centering
			\caption{SMEFT coefficients without flavour indices in \texttt{RGESolver}.}
			\label{tab:parnames0F}
			\begin{tabular}[t]{cc}
				\centering
				\begin{tabular}[t]{c|c}
					\multicolumn{2}{c}{Classes 1-3} \\ \hline 
					Coeff.     & \multicolumn{1}{c}{Name}      \\ \hline 
					$C_{G}$         & \texttt{CG}      \\
					$C_{\tilde{G}}$ & \texttt{CGtilde} \\
					$C_{W}$         & \texttt{CW}      \\
					$C_{\tilde{W}}$ & \texttt{CWtilde} \\
					$C_H$           & \texttt{CH}      \\
					$C_{H \Box} $   & \texttt{CHbox}   \\
					$C_{HD}$        & \texttt{CHD}    
				\end{tabular} & 
				\begin{tabular}[t]{c|c}
					\multicolumn{2}{c}{Class 4} \\ \hline 
					Coeff.    & \multicolumn{1}{c}{Name}         \\ \hline 
					$C_{HG}$         & \texttt{CHG}      \\
					$C_{H\tilde{G}}$ & \texttt{CHGtilde} \\
					$C_{HW}$         & \texttt{CHW}      \\
					$C_{H\tilde{W}}$ & \texttt{CHWtilde} \\
					$C_{HB}$         & \texttt{CHB}      \\
					$C_{H\tilde{B}}$ & \texttt{CHBtilde} \\
					$C_{HWB}$         & \texttt{CHWB}      \\
					$C_{H\tilde{W}B}$ & \texttt{CHWtildeB} 
				\end{tabular}
			\end{tabular}
			\vfill
		\end{minipage}
	\end{table}%

	\begin{table}[h!]
		\centering
		\scriptsize
		\renewcommand{\arraystretch}{1.3} 
		\caption{SMEFT coefficients with two flavour indices in \texttt{RGESolver}.}
		\label{tab:parnames2F}
		\begin{tabular}{ccc}
			\begin{tabular}[t]{c|c|c}
				\multicolumn{3}{c}{Class 5} \\ \hline 
				Coeff.   & Name   & Sym.      \\ \hline 
				$\mathrm{Re}(C_{eH})$ & \texttt{CeHR} & \multicolumn{1}{c}{WC1}   \\ 
				$\mathrm{Im}(C_{eH})$ & \texttt{CeHI}  & \multicolumn{1}{c}{WC1}\\ 
				$\mathrm{Re}(C_{uH})$ & \texttt{CuHR}  & \multicolumn{1}{c}{WC1}\\ 
				$\mathrm{Im}(C_{uH})$ & \texttt{CuHI}  & \multicolumn{1}{c}{WC1}\\ 
				$\mathrm{Re}(C_{dH})$ & \texttt{CdHR}  & \multicolumn{1}{c}{WC1}\\ 
				$\mathrm{Im}(C_{dH})$ & \texttt{CdHI}  & \multicolumn{1}{c}{WC1}
			\end{tabular} & 
			\begin{tabular}[t]{c|c|c}
				\multicolumn{3}{c}{Class 6} \\ \hline 
				Coeff.   & Name  & Sym. \\ \hline 
				$\mathrm{Re}(C_{eW})$ & \texttt{CeWR} & WC1 \\ 
				$\mathrm{Im}(C_{eW})$ & \texttt{CeWI} & WC1\\ 
				$\mathrm{Re}(C_{eB})$ & \texttt{CeBR} & WC1\\ 
				$\mathrm{Im}(C_{eB})$ & \texttt{CeBI} & WC1\\ 
				$\mathrm{Re}(C_{uG})$ & \texttt{CuGR} & WC1\\ 
				$\mathrm{Im}(C_{uG})$ & \texttt{CuGI} & WC1\\ 
				$\mathrm{Re}(C_{uW})$ & \texttt{CuWR} & WC1\\ 
				$\mathrm{Im}(C_{uW})$ & \texttt{CuWI} & WC1\\ 
				$\mathrm{Re}(C_{uB})$ & \texttt{CuBR} & WC1\\ 
				$\mathrm{Im}(C_{uB})$ & \texttt{CuBI} & WC1\\ 
				$\mathrm{Re}(C_{dG})$ & \texttt{CdGR} & WC1\\ 
				$\mathrm{Im}(C_{dG})$ & \texttt{CdGI} & WC1\\ 
				$\mathrm{Re}(C_{dW})$ & \texttt{CdWR} & WC1\\ 
				$\mathrm{Im}(C_{dW})$ & \texttt{CdWI} & WC1\\ 
				$\mathrm{Re}(C_{dB})$ & \texttt{CdBR} & WC1\\ 
				$\mathrm{Im}(C_{dB})$ & \texttt{CdBI} & WC1
			\end{tabular} &  
			\begin{tabular}[t]{c|c|c}
				\multicolumn{3}{c}{Class 7} \\ \hline 
				Coeff.     & Name & Sym.  \\ \hline
				$\mathrm{Re}(C_{Hl(1)})$ & \texttt{CHl1R}& WC2R \\ 
				$\mathrm{Im}(C_{Hl(1)})$ & \texttt{CHl1I} & WC2I\\ 
				$\mathrm{Re}(C_{Hl(3)})$ & \texttt{CHl3R} & WC2R\\ 
				$\mathrm{Im}(C_{Hl(3)})$ & \texttt{CHl3I} & WC2I\\ 
				$\mathrm{Re}(C_{He})$ & \texttt{CHeR} & WC2R\\ 
				$\mathrm{Im}(C_{He})$ & \texttt{CHeI} & WC2I\\ 
				$\mathrm{Re}(C_{Hq(1)})$ & \texttt{CHq1R} & WC2R\\ 
				$\mathrm{Im}(C_{Hq(1)})$ & \texttt{CHq1I} & WC2I\\ 
				$\mathrm{Re}(C_{Hq(3)})$ & \texttt{CHq3R}& WC2R \\ 
				$\mathrm{Im}(C_{Hq(3)})$ & \texttt{CHq3I} & WC2I\\ 
				$\mathrm{Re}(C_{Hu})$ & \texttt{CHuR} & WC2R\\ 
				$\mathrm{Im}(C_{Hu})$ & \texttt{CHuI} & WC2I\\ 
				$\mathrm{Re}(C_{Hd})$ & \texttt{CHdR} & WC2R\\ 
				$\mathrm{Im}(C_{Hd})$ & \texttt{CHdI} & WC2I\\
				$\mathrm{Re}(C_{Hud})$ & \texttt{CHudR} & WC1\\ 
				$\mathrm{Im}(C_{Hud})$ & \texttt{CHudI} & WC1  
			\end{tabular}
		\end{tabular}
	\end{table}

	\begin{table}[h]
		\centering
		\scriptsize
		\renewcommand{\arraystretch}{1.3} 
		\caption{SMEFT coefficients with four flavour indices in \texttt{RGESolver}.}
		\label{tab:parnames4F}
		\begin{tabular}[t]{ccc}
			\begin{tabular}[t]{c|c|c}
				\multicolumn{3}{c}{Class 8 $(\bar{L}L)(\bar{L}L)$} \\ \hline
				Coeff.     & Name &Sym. \\ \hline
				$\mathrm{Re}(C_{ll})$ & \texttt{CllR} & WC6R \\
				$\mathrm{Im}(C_{ll})$ & \texttt{CllI} & WC6I \\	
				$\mathrm{Re}(C_{qq(1)})$ & \texttt{Cqq1R} & WC6R \\
				$\mathrm{Im}(C_{qq(1)})$ & \texttt{Cqq1I} & WC6I \\	
				$\mathrm{Re}(C_{qq(3)})$ & \texttt{Cqq3R} & WC6R \\
				$\mathrm{Im}(C_{qq(3)})$ & \texttt{Cqq3I} & WC6I \\
				$\mathrm{Re}(C_{lq(1)})$ & \texttt{Clq1R} & WC7R \\
				$\mathrm{Im}(C_{lq(1)})$ & \texttt{Clq1I} & WC7I \\
				$\mathrm{Re}(C_{lq(3)})$ & \texttt{Clq3R} & WC7R \\
				$\mathrm{Im}(C_{lq(3)})$ & \texttt{Clq3I} & WC7I  \\ [7 pt]             
				\multicolumn{3}{c}{Class 8 $(\bar{L}R)(\bar{L}R)$} \\ \hline
				Coeff.     & Name & Sym.\\ \hline
				$\mathrm{Re}(C_{quqd(1)})$ & \texttt{Cquqd1R} & WC5 \\
				$\mathrm{Im}(C_{quqd(1)})$ & \texttt{Cquqd1I} & WC5 \\
				$\mathrm{Re}(C_{quqd(8)})$ & \texttt{Cquqd8R} & WC5 \\
				$\mathrm{Im}(C_{quqd(8)})$ & \texttt{Cquqd8I} & WC5 \\
				$\mathrm{Re}(C_{lequ(1)})$ & \texttt{Clequ1R} & WC5 \\
				$\mathrm{Im}(C_{lequ(1)})$ & \texttt{Clequ1I} & WC5 \\	
				$\mathrm{Re}(C_{lequ(3)})$ & \texttt{Clequ3R} & WC5 \\
				$\mathrm{Im}(C_{lequ(3)})$ & \texttt{Clequ3I} & WC5		
			\end{tabular} &
			\begin{tabular}[t]{c|c|c} 
				\multicolumn{3}{c}{Class 8 $(\bar{R}R)(\bar{R}R)$} \\ \hline
				Coeff.     & Name &Sym. \\ \hline
				$\mathrm{Re}(C_{ee})$ & \texttt{CeeR} & WC8R \\
				$\mathrm{Im}(C_{ee})$ & \texttt{CeeI} & WC8I \\
				$\mathrm{Re}(C_{uu})$ & \texttt{CuuR} & WC6R \\
				$\mathrm{Im}(C_{uu})$ & \texttt{CuuI} & WC6I \\
				$\mathrm{Re}(C_{dd})$ & \texttt{CddR} & WC6R \\
				$\mathrm{Im}(C_{dd})$ & \texttt{CddI} & WC6I \\	
				$\mathrm{Re}(C_{eu})$ & \texttt{CeuR} & WC7R \\
				$\mathrm{Im}(C_{eu})$ & \texttt{CeuI} & WC7I \\
				$\mathrm{Re}(C_{ed})$ & \texttt{CedR} & WC7R \\
				$\mathrm{Im}(C_{ed})$ & \texttt{CedI} & WC7I \\
				$\mathrm{Re}(C_{ud(1)})$ & \texttt{Cud1R} & WC7R \\
				$\mathrm{Im}(C_{ud(1)})$ & \texttt{Cud1I} & WC7I \\
				$\mathrm{Re}(C_{ud(8)})$ & \texttt{Cud8R} & WC7R \\
				$\mathrm{Im}(C_{ud(8)})$ & \texttt{Cud8I} & WC7I \\ [7 pt]             
				\multicolumn{3}{c}{Class 8 $(\bar{L}R)(\bar{R}L)$} \\ \hline
				Coeff.     & Name & Symmetry \\ \hline
				$\mathrm{Re}(C_{ledq})$ & \texttt{CledqR} & WC5 \\
				$\mathrm{Im}(C_{ledq})$ & \texttt{CledqI} & WC5 
			\end{tabular} & 
			\begin{tabular}[t]{c|c|c} 
				\multicolumn{3}{c}{Class 8 $(\bar{L}L)(\bar{R}R)$} \\ \hline
				Coeff.  & Name &Sym. \\ \hline
				$\mathrm{Re}(C_{le})$ & \texttt{CleR} & WC7R \\
				$\mathrm{Im}(C_{le})$ & \texttt{CleI} & WC7I \\
				$\mathrm{Re}(C_{lu})$ & \texttt{CluR} & WC7R \\
				$\mathrm{Im}(C_{lu})$ & \texttt{CluI} & WC7I \\
				$\mathrm{Re}(C_{ld})$ & \texttt{CldR} & WC7R \\
				$\mathrm{Im}(C_{ld})$ & \texttt{CldI} & WC7I \\
				$\mathrm{Re}(C_{qe})$ & \texttt{CqeR} & WC7R \\
				$\mathrm{Im}(C_{qe})$ & \texttt{CqeI} & WC7I \\
				$\mathrm{Re}(C_{qu(1)})$ & \texttt{Cqu1R} & WC7R \\
				$\mathrm{Im}(C_{qu(1)})$ & \texttt{Cqu1I} & WC7I \\
				$\mathrm{Re}(C_{qu(8)})$ & \texttt{Cqu8R} & WC7R \\
				$\mathrm{Im}(C_{qu(8)})$ & \texttt{Cqu8I} & WC7I \\
				$\mathrm{Re}(C_{qd(1)})$ & \texttt{Cqd1R} & WC7R \\
				$\mathrm{Im}(C_{qd(1)})$ & \texttt{Cqd1I} & WC7I \\
				$\mathrm{Re}(C_{qd(8)})$ & \texttt{Cqd8R} & WC7R \\
				$\mathrm{Im}(C_{qd(8)})$ & \texttt{Cqd8I} & WC7I \\
			\end{tabular} 
		\end{tabular}
	\end{table}

	Getters and setters take as first argument a string
	(\lstinline{name}), that identifies the corresponding parameter. The
	names that must be used to correctly invoke these functions are
	reported in
	Tables~\ref{tab:parnamesSM},~\ref{tab:parnames0F},~\ref{tab:parnames2F}
	and~\ref{tab:parnames4F}. The functions for non-scalar coefficients
	take as additional arguments the flavour indices ($2$ or $4$), that must be in the
	interval \lstinline{[0:2]}. 
	\begin{itemize}
		\item \method{void SetCoefficient(std::string name, double val)}:
		setter function for the scalar parameters. Sets the parameter
		\lstinline{name} equal to \lstinline{val}.
		\item \method{double GetCoefficient(std::string name)}: getter
		function for the scalar parameters. Returns the parameter
		\lstinline{name}.
		\item \method{void SetCoefficient(std::string name, double val, int i,
			int j)}: setter function for the parameters with two flavour
		indices. Sets the parameter \lstinline{name[i,j]} equal to
		\lstinline{val}.
		\item \method{double GetCoefficient(std::string name, int i, int j)}:
		getter function for the parameters with two flavour indices. Returns
		the parameter \lstinline{name[i,j]}.
		\item \method{void SetCoefficient(std::string name, double val, int i,
			int j, int k, int l)}: setter function for the parameters with
		four flavour indices. Sets the parameter \lstinline{name[i,j,k,l]}
		equal to \lstinline{val}.
		\item \method{double GetCoefficient(std::string name, int i, int j,
			int k, int l)}: getter function for the parameters with four
		flavour indices. Returns the parameter \lstinline{name[i,j,k,l]}.
		\item  \method{double GetCKMAngle(std::string name)}, \method{double GetCKMPhase()}: getter methods to access CKM matrix parameters. These methods should be called only after methods that choose a specific flavour basis (as \lstinline{GenerateSMInitialConditions()} or \lstinline{EvolveToBasis()}), in which case the CKM matrix is updated. 
		\item \method{double GetCKMRealPart(int i, int j)}, \method{double GetCKMImagPart(int i, int j)}: getter functions for the real and imaginary part of the $(i,j)$ element of the CKM matrix. These methods should be called only after methods that choose a specific flavour basis (as \lstinline{GenerateSMInitialConditions()} or \lstinline{EvolveToBasis()}), in which case the CKM matrix is updated. The indices must be in the range \lstinline{[0:2]}.
		\item \method{void SaveOutputFile(std::string filename, std::string
			format)}: saves the current values of the SMEFT parameters in the
		file \lstinline{filename}. Currently, the only available format is
		Susy Les Houches Accord \lstinline{"SLHA"} (\cite{Allanach:2008qq}).
	\end{itemize}

	\subsubsection*{Numerical integration parameters} 
	Here we list the methods related to the parameters that control the numerical evolution, as described in Section~\ref{subsec:numericalresolution}: 
	\begin{itemize}
		\item \method{void Setepsrel(double epsrel)} : sets $\epsilon_{\textrm{rel}}$ equal to \lstinline{epsrel} (the default value is $\epsilon_{\textrm{rel}}=5\times 10^{-3}$).
		
		\item \method{double epsrel()} : returns $\epsilon_{\textrm{rel}}$.
		
		\item \method{void Setepsabs(double epsabs)} : sets $\epsilon_{\textrm{abs}}$ equal to \lstinline{epsabs} (the default value is $\epsilon_{\textrm{abs}}=10^{-13}$).
		
		\item \method{double epsabs()} : returns $\epsilon_{\textrm{abs}}$.
	\end{itemize}

	\subsubsection*{General methods}
	\begin{itemize}
		\item \method{RGESolver()}: the default constructor.
		\item \method{$\thicksim$RGESolver()}: the default destructor.
		\item \method{Reset()}: resets all the SMEFT coefficients to 0 and the 
		SM parameters to their default value (in the up basis). 
		$\epsilon_{\textrm{abs}}$ and $\epsilon_{\textrm{rel}}$ are reset to 
		their default value. This function should be called after the evolution, if the same 
		instance of the class is used for another run.
	\end{itemize}

	\subsection{Writing and compiling a program using the library}
	\label{subsec:example}
	We discuss here the usage of the main methods of the class, in order to make the reader acquainted with \lstinline{RGESolver}'s functionalities. The examples discussed in this Section can be found in the \href{https://github.com/silvest/RGESolver/tree/main/Examples}{\lstinline{Examples} directory} of the \lstinline{GitHub} repository.\newline 
	Let us start with the simplest case: the renormalization group evolution from an high-energy scale $\Lambda=10000$ GeV to $\mu_{\mathrm{Low}} = 250$ GeV with a single non-zero SMEFT coefficient at the starting scale, namely $\coeff{HG}{}$.
	We do not report the whole program here, but we discuss just the crucial steps. This program is available under the name \lstinline{ExampleEvolution.cpp}. \newline 
	The library must be included with 
	\begin{lstlisting}
		#include "RGESolver.h"
	\end{lstlisting}
	Inside the \lstinline{main}, an instance of the class must be created: 
	\begin{lstlisting}
		RGESolver S;
	\end{lstlisting}
	We define the two energy scales (given in GeV) and we set the initial condition.
	\begin{lstlisting}
		double Lambda = 10000.;
		double muLow = 250.;	
		S.SetCoefficient("CHG",
		1./(Lambda*Lambda));
	\end{lstlisting}
	The (numeric) evolution is then performed using the line 
	\begin{lstlisting}
		S.Evolve("Numeric", Lambda, muLow);
	\end{lstlisting}
	At this point, the user can access the evolved coefficients via the
	\lstinline{GetCoefficient()} methods (see
	Tables~\ref{tab:parnamesSM},~\ref{tab:parnames0F},~\ref{tab:parnames2F}
	and~\ref{tab:parnames4F} for a list of all the names to be used). For
	example, this line prints on the terminal the evolved value of
	$\coeff{HG}{}$.
	\begin{lstlisting}
		std::cout<<"CHG("<<muLow<<"GeV):"<<S.GetCoefficient("CHG")<<std::endl;
	\end{lstlisting}
	The user should compile this program typing in the terminal:\footnote{Mac users might need to add the flag \lstinline{-std=c++11} (or greater).}
	\begin{lstlisting}
g++ -o app ExampleEvolution.cpp `rgesolver-config --cflags` `rgesolver-config --libs`
	\end{lstlisting}
	If the \lstinline{RGESolver} library is not in the predefined search path (as usually is the case for local installation), it may be necessary to specify the path needed for compilation and linking against \lstinline{RGESolver}. A \lstinline{rgesolver-config} script is available in the \lstinline{<CMAKE_INSTALL_PREFIX>}\lstinline{/bin} directory, which can be invoked with the following options:
	\begin{itemize}
		\item \verb!--cflags!: to obtain the include path needed for compilation against \lstinline{RGESolver}.
		\item \verb!--libs!:  to obtain the flags needed for linking against \lstinline{RGESolver}.
	\end{itemize} 
	A more advanced example can be found in \lstinline{ExampleBackRotation.cpp}. The goal is to perform the evolution from $\Lambda=10000$ GeV to $\mu_{\mathrm{Low}}=250$ GeV and perform the back-rotation.
	The default initial conditions for the SM parameters in Table~\ref{tab:SMinput} are given at the scale $\mu=173.6$ GeV. It is possible to generate the initial conditions for the SM parameters at any scale, both evolving the default input or using a different set of values. \newline 
	We thus define our custom input, starting from the energy scale at which the input is given:
	\begin{lstlisting}
		double SMinputScale = 91.0;
	\end{lstlisting}
	We then define the masses of the fermions (in GeV), the CKM parameters (the CKM phase must be expressed in radians), the gauge couplings, the quartic coupling and the Higgs' mass squared (in $\mathrm{GeV}^2$):
	\begin{lstlisting}
		double Muin[3] = {0.002, 1.2, 170.};
		double Mdin[3] = {0.006, 0.05, 5.2};
		double Mein[3] = {0.0005, 0.1, 1.2};
		double s12in = 0.225;
		double s13in = 0.003675;
		double s23in = 0.042;
		double deltain = 1.17;
		double g1in = .35;
		double g2in = .65;
		double g3in = 1.2;
		double lambdain = 0.14;
		double mh2in = 15625.;
	\end{lstlisting}
	The initial conditions at the scale $\Lambda$ are thus generated in the down basis through the solution of the pure SM RGE's (via numeric integration) simply using:
	\begin{lstlisting}
		S.GenerateSMInitialConditions(
		SMinputScale,Lambda,"DOWN","Numeric",g1in,g2in,g3in,lambdain,mh2in,
		Muin,Mdin,Mein,s12in,s13in,s23in,deltain);
	\end{lstlisting}
	As anticipated, the user can also take advantage of the default low-energy input 
	for the SM parameters via:
	\begin{lstlisting}
		S.GenerateSMInitialConditions(Lambda,"DOWN","Numeric");
	\end{lstlisting}
	Having generated the initial conditions for the SM parameters, the user can thus set the values of the SMEFT coefficients at the scale $\Lambda$ and perform the evolution to $\mu_{\mathrm{Low}}$, as done before. We don't discuss this since the syntax is the same of the previous case.\newline 
	The evolution and the back-rotation (in the down basis) are performed via: 
	\begin{lstlisting}
		S.EvolveToBasis("Numeric",Lambda,muLow,"DOWN");
	\end{lstlisting} 
	Since this method computes also the CKM matrix, we can access its parameters or its elements using the dedicated getter functions. For example, the following lines print on-screen $\sin \theta_{12}$ at the scale \lstinline{muLow}:
	\begin{lstlisting}	
		std::cout<<"Sin(theta12)("<<muLow<< " GeV): "<<
		S.GetCKMAngle("s12")<< std::endl;
	\end{lstlisting}
	
	\section{Execution times and comparison with \texttt{DSixTools}}
	\label{sec:efficiency}
	We tested our code against \lstinline{DSixTools}\,(\cite{dsixtools1,dsixtools2}), a \lstinline{Mathematica} package
	for the handling of the SMEFT (and the Low-energy Effective Field
	Theory, but we focus on the SMEFT part of the
	package). \lstinline{DSixTools}~solves the SMEFT RGE's numerically and via the
	approximation described in Section~\ref{subsec:resolution}, as \lstinline{RGESolver} does. We performed an
	evolution from $\Lambda$ to $\mu_{\textrm{Low}}=250$ GeV with both
	codes to compare the values of the coefficients after the
	evolution. The initial conditions at the high energy scale $\Lambda$
	for the SM parameters are obtained using pure SM RGE's to run them up
	to such scale. We set $\order{1/\Lambda^2}$ initial conditions for
	every SMEFT coefficient at $\mu=\Lambda$.\footnote{For SMEFT
		coefficients with flavour indices, we set $\ne 0$ at least one entry
		for each operator.} The result of this comparison is displayed in
	Table~\ref{tab:comparison}.\newline To compare the two results, we
	introduce
	$\Delta^{\mathrm{max}} = \max
	\lvert \coeff{i}{\mathtt{R}}(\mu_{\mathrm{Low}})-\coeff{i}{\mathtt{D}}(\mu_{\mathrm{Low}})\rvert/(1/\Lambda^2)$,
	with $\coeff{i}{\mathtt{R}(\mathtt{D})}(\mu_{\mathrm{Low}})$ the
	evolved coefficient computed by \lstinline{RGESolver}
	(\lstinline{DSixTools}). \newline We observe that the two codes produce the same
	results up to $\order{10^{-5}/\Lambda^2}$ ($\order{10^{-6}/\Lambda^2}$) in the case of numeric solution (approximate solution). Clearly, the agreement is
	better for the simpler solution, namely the approximate one. \newline
	\begin{table}[h!]
		\caption{Comparison between \lstinline{DSixTools}~and \lstinline{RGESolver}: the table on the left refers to the approximate solution, the one on the right to the numeric solution.} \label{tab:comparison}
		\centering
		\begin{tabular}{cc}
			\begin{tabular}{c|c}
				$\Delta^{\mathrm{max}}$ & $\Lambda $ (TeV)               \\ \hline
				$4.8 \times 10^{-6}$ & $1$\\ \hline
				$4.8 \times 10^{-6}$ & $10$\\ \hline 
				$5.1 \times 10^{-6}$ & $100$\\
			\end{tabular} 
			&  \hspace{10 pt}
			\begin{tabular}{c|c}
				$\Delta^{\mathrm{max}}$ & $\Lambda $ (TeV)               \\ \hline
				$1.2 \times 10^{-5}$ & $1$\\ \hline
				$1.2 \times 10^{-5}$ & $10$\\ \hline 
				$4.2 \times 10^{-5}$ & $100$\\
			\end{tabular} 
		\end{tabular}
	\end{table}
	
	Let us briefly discuss \lstinline{RGESolver}'s efficiency
	performances. We report in Table~\ref{tab:efficiency} the execution
	times that we measured (referring to a PC whose specifications are
	shown in Table~\ref{tab:specs}). The execution times consider not only
	the evolution, but also the input/output (setting the initial
	conditions for the SMEFT parameters and accessing them after the
	evolution).  These times should not be taken as exact, but as
	an order-of-magnitude estimate. \newline 
	This result should be compared to the execution times of \lstinline{DSixTools}, that can be up to $\order{20 \,\mathrm{s}}$ ($\order{10 \,\mathrm{s}}$) for the numeric (approximate) solution. This package performs a consistency check on the input that affects heavily the execution time for large inputs. Such impact is $\sim 8\,$s for the input used in the comparison between the two programs.\newline
	Conversely, considering just 3 non vanishing independent operators at $\mu = \Lambda$, the execution time is $\order{ 10\,\mathrm{s}}$ ($\order{2 \,\mathrm{s}}$) for the numeric (approximate) solution.
	
	\begin{table}[h!]
		\centering
		\def\arraystretch{1.3}
		\caption{Execution times of \lstinline{C++}~programs that use \lstinline{RGESolver}.} \label{tab:efficiency}
		\begin{tabular}{c|c}
			Operation                                                                                                            & Time                       \\ \hline
			\begin{tabular}[c]{@{}c@{}}approximate \\ solution\end{tabular}                                                      & $\order{2 \,\textrm{ms}}$  \\ \hline
			\begin{tabular}[c]{@{}c@{}}Numeric \\ solution\end{tabular}                                                          & $\order{50\,\textrm{ms}}$  \\ \hline
			\begin{tabular}[c]{@{}c@{}}Numeric solution \\ + generation of SM initial conditions \\ + back-rotation\end{tabular} & $\order{100\,\textrm{ms}}$ \\ 
		\end{tabular}
	\end{table}
	
	\begin{table}[h!]
		\def\arraystretch{1.3}
		\caption{Technical specifications of the computer used for the test.} \label{tab:specs}
		\centering
		\begin{tabular}{c|c}
			CPU & Intel$^{\textrm{\textcopyright}}$ Core™ i7-6500U CPU $@$2.50GHz×4 \\ \hline
			RAM & 7,7 GiB                                                             \\ \hline
			OS  & Ubuntu 20.04.2 LTS           \\                                   
		\end{tabular}
	\end{table}

	\section{Conclusions}
	\label{sec:conclusions}
	\lstinline{RGESolver} is an open-source \lstinline{C++}~library that performs the
	renormalization group evolution in the context of the SMEFT at
	dimension-six level, in the most general flavour scenario. Only
	operators that preserve baryon and lepton number are considered.
	\lstinline{RGESolver} was designed to be easy to use, highly efficient
	and suitable to perform extensive phenomenological analysis. To this
	aim, it will be included in \lstinline{HEPfit}~(\cite{hepfit}), a multipurpose
	and flexible analysis framework that can be used for fitting models to
	experimental and theoretical constraints. \lstinline{RGESolver}
	outperforms \texttt{DSixTools}~in execution time by two orders of magnitude,
	while retaining an excellent accuracy. Further details and the
	full documentation can be found on
	\href{https://github.com/silvest/RGESolver}{\texttt{GitHub}}.

	\section*{Acknowledgments}
	
	S.D.N. would like to thank his colleagues (and
	friends) from Rome and Padua for their contributions to early tests of \texttt{RGESolver} and A. Vicente, whose help was essential to complete the comparison between \texttt{RGESolver} and \texttt{DSixTools}. This work was
	supported in part by the Italian Ministry of Research (MIUR) under
	grant PRIN 20172LNEEZ. The Feynman diagrams shown in this work
	were drawn with \texttt{TikZ-Feynman} (\cite{tikz-feynman}). 
	
\section*{Data Availibility Statement} 
	The manuscript has associated data in a data repository. [Authors’ comment: The code release described in this manuscript is available at \href{https://doi.org/10.5281/zenodo.8006774
	}{https://doi.org/10.5281/zenodo.8006774}].

		
		
		
		

	\bibliographystyle{JHEP}
	\bibliography{Biblio1}

\end{document}

%% file: AdditionalPreamble.tex
\usepackage{csquotes} 

\usepackage{slashed} 
\usepackage{subcaption} 

\usepackage{tikz}
\usetikzlibrary{shapes.geometric}
\usepackage[compat=1.1.0]{tikz-feynman} 

\usepackage{comment}
\usepackage{alltt}
\usepackage{fancyvrb}

\usepackage{braket}
\usepackage{amssymb, amsmath, mathtools,slashed, breqn, amsthm, mathrsfs, esint, bm}

\usepackage{listings}

\usepackage{float}

%% file: CustomCommands.tex
\newcommand{\op}[2]{ \mathcal{O}_{#1} ^{#2} } 

\newcommand{\coeff}[2]{ \mathcal{C}_{#1} ^{#2} }
\DeclareFontFamily{OT1}{pzc}{}
\DeclareFontShape{OT1}{pzc}{m}{it}{<-> s * [1.10] pzcmi7t}{}
\DeclareMathAlphabet{\mathpzc}{OT1}{pzc}{m}{it}

\newcommand{\group}[1]{\mathrm{#1}}
\newcommand{\repr}[1]{ \mathbf{#1} } 
\newcommand{\qm}[3]{ (#1,#2)_{#3} } 

\newcommand{\adm}{ \Gamma}

\newcommand{\hy}[1]{ \mathsf{y}_{#1}}
\newcommand{\yuk}[1]{{Y}_{#1}}
\newcommand{\gl}[2]{ T_{#1} ^{#2}}
\newcommand{\pauli}[2]{ \tau_{#1} ^{#2}}

\newcommand{\order}[1] {\mathcal{O }\left( #1 \right)}

\newcommand{\lag}{ \mathcal{L}} 

\newcommand{\method}[1]{{\texttt{#1}}}

\makeatletter
\newcommand{\overleftrightsmallarrow}{\mathpalette{\overarrowsmall@\leftrightarrowfill@}}
\newcommand{\overrightsmallarrow}{\mathpalette{\overarrowsmall@\rightarrowfill@}}
\newcommand{\overleftsmallarrow}{\mathpalette{\overarrowsmall@\leftarrowfill@}}
\newcommand{\overarrowsmall@}[3]{%
	\vbox{%
		\ialign{%
			##\crcr
			#1{\smaller@style{#2}}\crcr
			\noalign{\nointerlineskip}%
			$\m@th\hfil#2#3\hfil$\crcr
		}%
	}%
}
\def\smaller@style#1{%
	\ifx#1\displaystyle\scriptstyle\else
	\ifx#1\textstyle\scriptstyle\else
	\scriptscriptstyle
	\fi
	\fi
}
\makeatother
\newcommand{\lrarrow}[1]{\overleftrightsmallarrow{#1}}